\documentstyle[twocolumn]{mn}

\title{ Magnetic Flux Expulsion in the Powerful Superbubble Explosions and 
the $\alpha{\mbox -}\Omega$ Dynamo.}

\author[R.R.~Rafikov and R.M.~Kulsrud]
       {R.R.~Rafikov$^1$ and R.M.~Kulsrud$^1$ \\
           $^1$ Peyton Hall, Princeton University, Princeton, NJ, 08544, USA}
\date{Accepted 2000  .
     Received  1999 ;
     in original form 1999}

\pagerange{\pageref{firstpage}--\pageref{lastpage}}

\pubyear{2000}

\begin{document}

\maketitle

\label{firstpage}

\begin{abstract}
The possibility of the magnetic flux expulsion from the Galaxy in the 
superbubble (SB) explosions, important for the $\alpha{\mbox -}\Omega$
dynamo,
 is considered. Special emphasis 
is put on the investigation of the  downsliding of the matter from the top of 
the shell formed by the SB explosion which is able to influence the
kinematics of the shell. It is shown that either Galactic gravity
or the development of the Rayleigh-Taylor instabilities in the shell,
limit the SB expansion, thus, making impossible magnetic flux 
expulsion. The effect 
of the cosmic rays in the shell on the sliding is considered
 and it is shown that it is negligible compared to Galactic gravity.
Thus, the question of possible mechanism of flux expulsion 
in the $\alpha{\mbox -}\Omega$ dynamo remains open.
\end{abstract}


\section{Introduction}

The magnetic field in our Galaxy and in other spiral galaxies is
usually believed to have been amplified from a weak seed field
by a hydromagnetic dynamo,
which exists due to the presence  of the large-scale differential rotation 
and small-scale cyclonic turbulence in the Galaxy (Parker 1970, 1971;
Vainshtein $\&$ Ruzmaikin 1971, 1972; Moffat 1978). 
It has been  suggested that
 any primordial magnetic field
could be expelled from the Galaxy by the dynamic motions in less than 
a billion years (Parker 1971) so it would seem that some flux amplification
is necessary to explain the Galactic field. 
The theory of such a dynamo 
has been formulated in a precise way through the mean field equations 
and the solutions to these equations indicate that the field would 
be amplified. The resulting 
field appears to correspond to the magnetic field patterns in our galaxy 
and others.    

On the other hand, a number of criticisms of this theory have emerged.
One of them concerns the intense development of the small scale fields,
which could  damp the turbulence and stop the dynamo action unless
they saturate at levels which do not interfere with mean field dynamo 
(Kulsrud $\&$ Anderson 1992; Parker 1992; Vainshtein $\&$
Cattaneo 1992). 

Another unresolved problem
is the expulsion of flux from the galactic disc.  
A very important point is that the theory of the $\alpha{\mbox -}\Omega$
dynamo predicts the amplification of some
small preexisting magnetic field only if some magnetic diffusion is present
 in the Galaxy. However,
too large a diffusion is destructive for the dynamo, because
of its dissipative role in the process of generation and it seems attractive
to suppose that a dynamo without diffusion at all will be the most effective.
But Ruzmaikin, Shukurov, $\&$ Sokoloff (1988) showed that this is not possible.
The physical reason for this is rooted in the very strong flux freezing 
of the galactic plasma, because magnetic lines cannot break
and the number of field lines in the disc can be increased 
only by toroidal stretching, which
is accomplished by the dynamo action. Any stretching creates field
of both signs, to conserve the total flux, and, for net amplification to occur,
those portions of field lines which are of the wrong sign must be 
expelled from the Galaxy. In the standard $\alpha{\mbox -}\Omega$
dynamo theory this is done by magnetic diffusion.   

Ruzmaikin, Shukurov, $\&$ Sokoloff (1988) 
considered  $\alpha{\mbox -}\Omega$-dynamo in the case of 
the thin disc and
demonstrated that temporal evolution of the magnetic flux is governed 
by the following set of equations:   
\begin{equation}
\frac{\partial}{\partial t}\int\limits^1_0 B_r(t,z)dz=\beta
\left.\frac{\partial B_r}{\partial z}\right|^1_0,
\label{flux1}
\end{equation}
\begin{equation}
\frac{\partial}{\partial t}\int\limits^1_0 B_{\varphi}(t,z)dz=\beta
\left.\frac{\partial B_{\varphi}}{\partial z}\right|_0^1+G\int\limits^1_0 
B_r dz,
\label{flux2}
\end{equation}
where $G=r\partial \Omega/\partial r$-measure of the differential rotation,
$\beta$-magnetic diffusion, and $0$ and $1$ correspond to the center plane and
boundary of the Galactic disc.

The $\beta$ terms represent the expulsion of flux.
It is clear, that setting $\beta=0$ in (\ref{flux1}) and (\ref{flux2}) we 
immediately get that
\begin{equation}
\int\limits^1_0 B_r dz=const,
\label{flux3}
\end{equation}
and that $\int\limits^1_0 B_{\phi}dz$ may grow only linearly due to the 
stretching of lines 
 by the galactic differential rotation. Thus, there is no 
exponential field growth, essential to the dynamo. 
One can easily see that 
for the dynamo to operate, there must be some  nonzero   flux escape
 through the upper boundary of the disc, 
that is $\beta\neq0$.

This is consistent with the topological constraint that the total 
number of lines of force including those negative lines expelled from 
the disc must be constant. Indeed, a simple estimate of the expulsion 
terms making use of the numerical results in Ruzmaikin, 
Shukurov,
 $\&$ Sokoloff (1988) shows that the negative flux expelled during one 
growth cycle is comparable to the positive flux in the disc at the beginning 
of the  e-folding.

Although the $\alpha{\mbox -}\Omega$
dynamo theory is complicated, this physical intuition of flux expulsion 
can be considered in the absence of these complications. Further, the
mechanism of expulsion need not be tied to the
$\beta$ diffusion inside the disc. 

The main problem with the expulsion of flux is that this flux is loaded 
with matter so that it is related
 to the expulsion of matter against the strong 
gravity of the galactic disc. The most likely process to expel flux is 
the phenomenon of sequential supernova (SN) or superbubbles (SB), 
which sweep up matter into dense, radiatively cooled shells.
Magnetic field, tied to the matter due to the strong flux freezing
in the ISM, is also swept up and deposited in these shells. 
If some part of the shell leaves the Galaxy, it carries the frozen-in
magnetic field with it, thus producing the flux expulsion.  

But most of 
these superbubbles  are not powerful enough to expel matter out 
of the gravitational well of the disc. The only possibility for flux 
expulsion seems to be:
 as the  bubble expands, 
the field lines in the shells
of SBs are not horizontal but form arcs, along which 
matter can slide down, lowering the amount of matter on the top
of the lines and allowing some flux to escape.  

There is further difficulty with the mechanism involving the SBs 
which is relevant to its application to the $\alpha{\mbox -}
\Omega$-dynamo. The $\alpha{\mbox -}
\Omega$-dynamo assumes 
{\it small-scale} turbulence while the cavities produced by SB explosions 
may  reach $\sim 500$ pc or larger, 
which is  greater than some of the length scales of the galactic disk. 
For the dynamo theory in its conventional form
to be applicable it is important
that turbulence be small scale, because it involves
 the expansion of  the turbulent electromotive force ${\cal E}$,
which describes the effect of turbulent motions on the mean (or ensemble-
averaged) magnetic field, in terms of the mean magnetic field itself 
and its spatial derivatives:
\begin{equation}
{\cal E}_i=\alpha_{ij}<B_j>+\beta_{ijk}\frac{\partial <B_j>}{\partial x_k}
\end{equation}  
(Moffat 1978)

If the scale of the turbulence is too large, then the 
 expansion is invalid  and usual
$\alpha{\mbox -}\Omega$-dynamo theory must be modified.  
That's why, for example, direct application of  $\alpha$ and $\beta$ tensors
calculated by Ferri$\grave {\rm e}$re (1995, 1998)
for SBs and SNs to the $\alpha{\mbox -}\Omega$-dynamo theory
can lead to an overly optimistic estimates of the rate of flux escape.

Although $\alpha{\mbox -}\Omega$-dynamo theory is not strictly applicable 
to the case of SBs, the actual operation of them in amplifying the field is
clear from the work of Ferri$\grave {\rm e}$re (1991, 1995, 1998). It
is also clear that the rapid escape of the flux from the disk is essential.
In this paper we show that because of the deep gravitational well of the 
Galaxy it is difficult for the matter and field lines to escape, and
consequently for the mean field to grow.
In Ferri$\grave {\rm e}$re's works she finds the lines of force rising 
with the SB but does not follow them long enough to see that they must fall
back into the disk and inhibit the growth of the field.

In this note we quantitatively examine the dynamics of the rising  
field lines carried by SBs and show that even 
with sliding the matter and flux are unlikely to escape.
Thus, the requirement of the escape of flux provides a strong constraint for
the $\alpha{\mbox -}\Omega$
dynamo to overcome if it is to amplify the galactic magnetic field.

\section{Sliding of matter from the top of SB; formulation of the problem}

Multiple supernovae from OB associations can carve out large cavities of 
hot gas, called superbubbles. McCray $\&$ Snow (1979) first described them. 
When the SB expands into surrounding medium it sweeps up interstellar
matter, giving rise to a massive expanding
shell. Inside the volume surrounded by this 
shell a hot rarefied
low-density gas is contained which provides the pressure  driving 
further expansion of the shell. 
The energetic source for sustaining this pressure
is provided by continuous energy input from the SN explosions in 
the center of SB. Weaver et al. (1977) 
calculated the evolution of the
bubble driven by the continuous wind from the central source and 
Mac Low $\&$ McCray (1988) applied this theory to the case of supernovae 
driven SBs. They show that the radius of such a SB, expanding in
a uniform medium of density $\rho_0$ with continuous energy input in the 
center 
$L_{SN}=L_{38} 10^{38}~{\rm ergs~ s^{-1}}$ (the luminosity of 
SB conveniently expressed 
in units of
$10^{38}~{\rm ergs~ s^{-1}}$), is given by
\begin{equation}
R(t)=\left(\frac{125}{154\pi}\right)^{1/5}L_{SN}^{1/5}
\rho_0^{-1/5}t^{3/5}=267\left(\frac{L_{38}t_7^3}{n_0}\right)^{1/5}{\rm pc},
\label{R}
\end{equation}
where $t_7=t/10^7$ yr,
with the velocity of the envelope changing as
\begin{equation}
u(t)=\dot R(t)\approx 15.7\left(\frac{L_{38}}{n_0 t_7^2}
\right)^{1/5}~{\rm km~ s^{-1}}.
\end{equation}
The inner pressure in the volume bounded by the shell varies as
\begin{eqnarray}
P_{in}(t)=\frac{7}{(3850\pi)^{2/5}}L_{SN}^{2/5}
\rho_0^{3/5}t^{-4/5}\nonumber\\
=4.1\times 10^{-12}
\left(\frac{L_{38}^2 n_0^3}{t_7^4}\right)^{1/5}~{\rm ergs~sm^{-3}},
\label{pressure}
\end{eqnarray}
due to the work done on the expansion of the shell and the energy 
injection in the center of SB.

When the shell expands in the real Galactic environment, there is also a 
gravitational force which tends to slow down the 
vertical expansion. Also, the 
distribution of medium is highly inhomogeneous with height $z$ over the 
Galactic plane. Various components of ISM have different length scales 
and characteristic densities, but they all have exponential or Gaussian 
decreasing profiles in $z$ and thus drop very rapidly with height. 
This effect is very noticeable
for powerful superbubbles, for which the expansion radius may exceed the
height
scale of the matter distribution. As we will see 
this can lead to Rayleigh-Taylor
instabilities. 

At the final stage of the SB expansion, its velocity becomes comparable to the 
sound velocity of the ambient ISM and shock wave will no longer
exist. However, this does not prevent the SB from further expansion, because
there is still a lot of momentum in its massive shell of swept up material
and the ram pressure at high $z$ is negligible.
 This means that expansion actually continues until the velocity of the shell
drops to zero under the action of Galactic gravity:
\begin{equation}
u=0.
\label{ucs}
\end{equation}
 This stopping condition is different from the one which 
Ferri$\grave {\rm e}$re used 
(1998) in her investigation  of the role of SBs in the 
$\alpha-\Omega$ dynamo. Her condition (expansion stops when the shell velocity
is of the order of speed of sound in the ambient ISM, in other
words when the pressure inside the SB becomes comparable to the 
ram pressure) is applicable only for
massless shells without any inertia. In reality shell is quite massive
and the momentum stored in it drives further expansion of the bubble
against the ram pressure and the galactic gravity.

Also when the SB expands in the presence of the gravity, 
the swept-up matter in its
shell is likely to slide downward along the shell
as mentioned before. 
This can influence the  efficiency of flux
removal by SBs. Obviously, in the absence of sliding, escaping matter takes 
with it all the frozen-in magnetic flux. Now, if we allow the matter in 
the shell to slide downward perpendicular to the field $B$ in the shell
it will take the magnetic field lines away from the 
top of the SB, 
thus reducing the magnetic flux to be removed. But it can also 
have a significant effect on the dynamics of the SB itself, because 
as the matter slides from the top in any direction, 
the upper part of the shell becomes
lighter and acceleration due to the inner pressure of the hot 
gas will gets larger in proportion to the surface density decrease, while
the gravitational deceleration  
stays the same. This results  in some additional 
acceleration of the top of the shell and, in principle, this can significantly 
change the dynamics of this part of SB if there is  enough time before 
the shell stops. If this happens, and a substantial part 
of the mass slides down, the top of the shell may continue its expansion
upward and drive the remaining matter and frozen-in flux to a further 
distance from the galactic plane and possibly expel the flux entirely
from the Galaxy. 
For this reason it is very important to 
estimate the numerical value of this effect. 

\section{Basic equations}

Giuliani (1982) gave a general formulation of the thin-shell approximation for 
hypersonic, hydromagnetic flows, axisymmetric about the $z$ axis, 
including motions along the shell.
 We use his equations 
 in our description of gravitational matter removal from the top of SB.
All our further consideration is restricted to the case when the 
angular distance of the shell element from the shell top, $\theta$,
 is very small,
$\theta\ll 1$.

We  suppose the expansion of the SB 
 is described by some  law $R=R(t)$. 
We will also suppose for simplicity that the 
form of the shell near its top can be roughly approximated by a sphere.
The validity of this assumption will be discussed later.

We made some further simplifications, one of which is the neglect of the 
pressure gradient along the shell. It is clear enough that any such
pressure gradient  would only reduce the downsliding 
of the matter. This means that our estimate is
only an upper limit of the sliding and the 
real sliding will actually be smaller. 

We also completely neglect the influence of the magnetic field 
on the dynamics of the sliding. This assumption is justified in the early
periods of the Galaxy's life, when the magnetic field was weak.
At the present time this 
is not completely valid, because the magnetic field is 
strong and magnetic tension may play some dynamical role in the expansion 
process. SBs with strong magnetic field were considered analytically by  
Ferri$\grave {\rm e}$re
(1991) and numerically by Tomisaka (1992).

Thus, the only external force in our analysis is the gravity
due to the stars and ISM in the Galaxy. We take gravity
as given, because the self-gravity 
of the bubble is negligible.
With this in mind the system of the equations of Giuliani describing the 
gravitational fall of matter from the top of the shell reduce to
\begin{equation}
\frac{\partial}{\partial t}(R^2\sin\theta\sigma)=\rho_0 u R^2\sin\theta - 
\frac{\partial}{\partial \theta}\left(R(t)\sin\theta\sigma v_{\parallel}
\right), 
\label{cont}
\end{equation}
\begin{equation}
\frac{\partial v_{\parallel}}{\partial t}+
\frac{\rho_0 u v_{\parallel}}{\sigma}+
\frac{v_{\parallel}}{R}\left(u+
\frac{\partial v_{\parallel}}{\partial \theta}\right)- g\sin\theta=0. 
\label{motion}
\end{equation}

Here $v_{\parallel}$ is the tangential velocity of the matter along the 
shell arising from the presence of gravity, $u=\dot R (t)$ is the 
velocity of the shell expansion (dot means time derivative), 
$\sigma$ is the surface density of the shell,
$\rho_0$ is the density of the unperturbed gas in front of the shell, and
$g$ is the gravitational acceleration. 

Now, at an early stage of SB expansion, when its radial velocity is very high
and the transverse velocity due to the galactic gravity is not very large,
the effect of sliding is negligible, because the total velocity of the 
shell element is directed almost radially. Thus, it is reasonable 
to suppose that the
 most noticeable effect of sliding will occur only during the 
later stages of the shell evolution, when it sufficiently slows down in 
the radial 
direction.  But by the onset of this later stage the SB has expanded
in the direction perpendicular to the galactic plane to a distance 
larger than the scale height of matter distribution. This reasoning 
allows us to neglect the second
term proportional the 
ambient density in equation (\ref{motion}) . It makes equation (\ref{motion})
completely independent of equation (\ref{cont}) since it then
contains no terms in  $\sigma$.

However, in equation  (\ref{cont}), we must keep terms in $\rho_0$
because $\sigma$ still grows due to swept up matter.

Also we can neglect
the term $\partial v_{\parallel}/\partial\theta$
 in (\ref{motion})
compared to the radial velocity $u$. We discuss
this omission later. Thus equations (\ref{cont}) and (\ref{motion}) 
reduce to the following
simplified  system 
of equations:
\begin{eqnarray}
\frac{\partial}{\partial t}(R^2(t)\sin\theta \sigma)=
\rho_0 u R^2(t)\sin\theta
\nonumber\\ 
-R(t)\sigma\frac{\partial}{\partial \theta}\left(\sin\theta v_{\parallel}
\right)-
  R(t)\sin\theta v_{\parallel}\frac{\partial \sigma }{\partial \theta}, 
\label{cont-si}
\end{eqnarray}
\begin{equation}
\frac{\partial v_{\parallel}}{\partial t}+
v_{\parallel}
\frac{u}{R(t)}=
g\sin\theta. 
\label{m_si}
\end{equation}

In general, we assume that gravitational acceleration, $g$, is a function 
of the  $z$-coordinate in the Galaxy.
Bearing in mind that $u=\dot R(t)$ and $v_{\parallel}=0$ at $t=0$, we can 
integrate equation (\ref{m_si}) to get
\begin{eqnarray}
v_{\parallel}(t)=\frac{\sin\theta}{R(t)}\int\limits^{t}_{0} 
R(t^{\prime})g(t^{\prime})dt^{\prime}
\nonumber\\
=\frac{\sin\theta}{R(t)}\int\limits^{t}_{0} 
R(t^{\prime})g(z_0+R(t^{\prime}))dt^{\prime},
\label{v_par}
\end{eqnarray}
where $z_0$ is the height in the Galactic plane where the 
explosion occured. 

We see that $v_{\parallel}\propto \sin\theta$. This means that in
equation (\ref{cont-si}) we may neglect the last term near the top of SB
since it is proportional to $\sin^2\theta$. Then equation (\ref{cont-si})
further reduces to 
\begin{eqnarray}
\dot\sigma+
\frac{1}{R}
\left[2\dot R+\frac{1}{\sin\theta}\frac{\partial}{\partial \theta}
\left(\sin\theta v_{\parallel}\right)\right]\sigma=
\rho_0 \dot R. 
\label{c-si}
\end{eqnarray}
This equation  can be integrated with the
initial condition $\sigma=0$ at $t=0$:
\begin{eqnarray}
\sigma(t)=\frac{1}{R^2(t)}\exp\left(-\int\limits^{t}_{0}\kappa(t^{\prime})
dt^{\prime}\right) \nonumber\\
\times
\int\limits^{t}_{0}R^2 \dot R \rho(z_0+R(t^{\prime}))
\exp\left(\int\limits^{t^{\prime}}_{0}\kappa(t^{\prime\prime})
dt^{\prime\prime}\right)
dt^{\prime},
\label{sigma}
\end{eqnarray}
where
\begin{equation}
\kappa=\frac{1}{\sin\theta}\frac{\partial}{\partial \theta}
\left(\sin\theta v_{\parallel}\right).
\label{kappa}
\end{equation}

Substitution of expression (\ref{v_par}) into (\ref{kappa})
gives 
\begin{eqnarray}
\kappa=2\cos\theta \frac{1}{R^2(t)}\int\limits^{t}_{0} 
R(t^{\prime})g(t^{\prime})dt^{\prime}
\nonumber\\
\approx
\frac{2}{R^2(t)}\int\limits^{t}_{0} 
R(t^{\prime})g(z_0+R(t^{\prime}))dt^{\prime}, 
\label{kappa-s}
\end{eqnarray}
near the top of SB.
    
	From the formulae (\ref{sigma}) and (\ref{kappa}) it is easy
to see that the importance of sliding near the top is determined by the 
quantity
\begin{equation}
\zeta=e^{\kappa}=\exp\left(
\frac{2}{R^2(t)}\int\limits^{t}_{0} 
R(t^{\prime})g(t^{\prime})dt^{\prime}\right), 
\label{lambda}
\end{equation}
which can be determined for any given $R(t)$.

If $\zeta\simeq 1$, we can safely neglect the sliding of matter but
if $\zeta\gg 1$, sliding will play important role in the 
dynamics of late stages of SB expansion.

\section{Influence of sliding on the shell expansion}

 Let us examine the equation for $R(t)$ near the top of the bubble taking 
into account  the effect
of the unloading the matter from the shell's top on the radial 
expansion of the SB itself.

\begin{figure}
\vspace{16.0cm}
\includegraphics{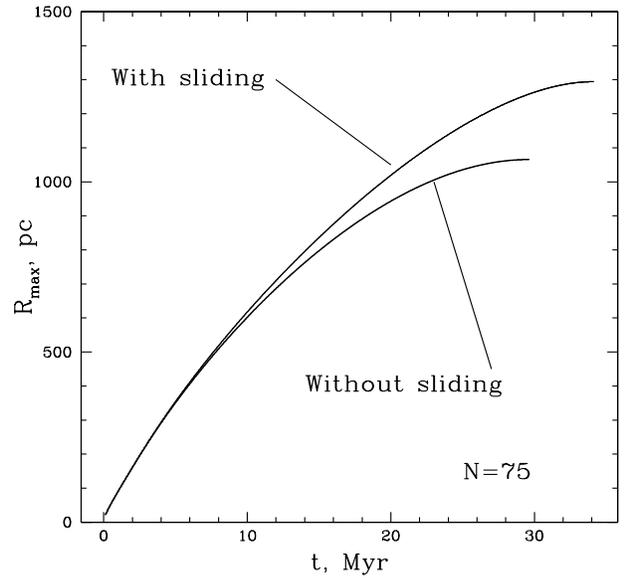}
\includegraphics{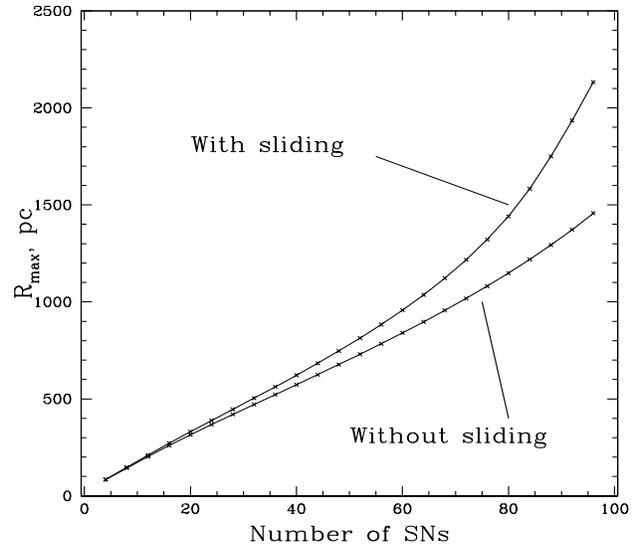}
\caption{
(a) Dependence of polar radius $R$ (at $\theta=0$) upon the expansion 
time $t$ for the case of the sliding of matter due to the Galactic gravity
(upper curve) and without sliding (lower curve). The case of 75 SNs in
SB is considered, corresponding to the luminosity $L=7.5 \times 
10^{37}$ ergs s$^{-1}$.  
Note that SB with sliding 
has larger final size, than that without sliding.
(b) Dependence of the maximum radii of the SB, $R_{max}$, 
on the luminosity (expressed through the number
of SNs in the SB in equation (\ref{lum})) with (upper curve) and
without sliding(lower curve). The difference in final sizes of the SBs
with and without this effect 
increases with the bubble's luminosity $L$.}
\label{sliding}
\end{figure}

To do this we
 consider a solid angle $d\Omega$ of the shell near its top. We can 
write the following equation for its motion:
\begin{equation}
\frac{d}{dt}\left(\sigma \dot R R^2 d\Omega\right)=\left(P_{in}-P_{out}\right)
R^2 d\Omega -R^2\sigma g d\Omega.
\label{motion1}
\end{equation}

We can combine this equation with equation (\ref{sigma}) 
to find $R(t)$ and $\sigma(t)$ near $\theta=0$
 as a functions of time $t$ if we know how $P_{in}$ behaves. 
All other quantities, namely $P_{out}$ and $g$ are given empirically, 
as a function of $z$,
by the position of 
the shell's top.

Equations (\ref{motion1}) and (\ref{c-si}) can be combined to give
\begin{equation}
\ddot R+\dot R^2\frac{\rho_0}{\sigma}-\frac{2\dot R}{R^2}
\int\limits_0^t R(t^{\prime})g(t^{\prime})dt^{\prime}=
\frac{P_{in}-P_{out}}{\sigma}-g
\label{rexpa}
\end{equation}
and together with equation (\ref{sigma}) give $R(t)$.

These equations are solved numerically to get the behavior of $R$. 
First we supposed that the  inner pressure is governed by  
equation (\ref{pressure}). This underestimates the pressure and the 
height reached. But if there is no escape of flux under this assumption
there is certainly no escape in real conditions.   
 We have taken 
the model of ISM from  
Ferri$\grave {\rm e}$re (1998), that is we supposed that density and pressure
of ISM are contributed by $5$ components, having different number densities,
length scales and temperatures: neutral, cold, warm, ionized, and hot.
    We also use her approximation for the gravitational
acceleration $g$.  
  We suppose for simplicity that the luminosity   of the SB is constant 
in time during 37 Myr, until the death of the $8 M_{\odot}$ stars
(Ferri$\grave {\rm e}$re 1995),
 and equal to 
\begin{equation}
L=10^{36}\times N~{\rm erg~s^{-1}},
\label{lum}
\end{equation}
 where
$N$ is the number of SNs in the star cluster. For this calculation 
we suppose that the inner pressure $P_{in}$ changes in accordance with
equation (\ref{pressure}) during the first $37$ Myr of the SB expansion,
with $\rho_0$  the mass density at the site of explosion.
After $37$ Myr the interior cools adiabatically because 
the inner pressure and inertia of the shell continue
 to drive the shell expansion so that the volume bounded by the 
shell 
increases. 
We also carry out  calculations for the different pressure law.
But we  neglect 
radiative cooling of the hot gas 
 in the interior of the bubble during the entire explosion, so that real 
SB expansion is always smaller than we obtain here.

In Figure \ref{sliding}a we show the dependence of 
$R$ upon time $t$ for the  case of a SB with $N=75$ SNs in it, going off
near the Sun at an initial Galactic altitude $100$ pc. For comparison we
also depict the curve without sliding for the same SB, which is obtained by
setting $g$ in the integral in the left-hand side of equation 
(\ref{rexpa}) to zero
(but not in the right-hand side!). 

If the sliding is taken into account, the time when expansion 
stops is $t_s=35$ Myr and $R_{max}=1336$ pc; 
neglecting
sliding we get $t_s=30$ Myr and  $R_{max}=1094$ pc.  
It is obvious that effect of sliding should be more 
pronounced in more powerful bubbles because of their longer lifetimes and 
stronger gravity at the heights to which they bring the matter. To illustrate 
this we plot in Figure \ref{sliding}b the maximum 
radii of SBs of various luminosities
with and without sliding for the same conditions 
as the Figure \ref{sliding}a. It is
clearly seen that sliding plays important role for powerful SBs, 
making their final radii dozens of percent larger than that without sliding. 

\begin{figure}
\vspace{8.0cm}
\includegraphics{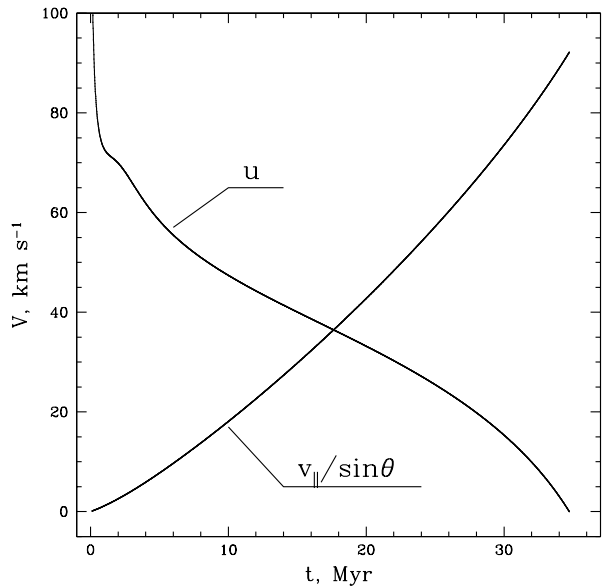}
\caption{
Plots of the expansion velocity of the SB top  $u$ (declining curve) and
the velocity of the downsliding of the matter
along the shell $v_{\parallel}$ (growing curve), divided by 
$\sin\theta$. These plots
 show that at the early stages of the shell expansion
we can neglect $\partial v_{\parallel}/\partial \theta$ compared
to $u$ in equation (\ref{motion}).}
\label{velocities}
\end{figure}

In Figure \ref{velocities} we plot,
for comparison, the expansion velocity of the $N=75$ SB top  and
the velocity of sliding along the shell for the same SB, divided by 
$\sin\theta$ (which virtually equals to  $\partial v_{\parallel}/\partial
\theta$ near the SB top). One can see that, up to the 
first $20$ Myr, $u$ is larger 
than $\partial v_{\parallel}/\partial \theta$, which justifies our 
neglect of corresponding term in equation (\ref{motion}). By that
time most of the SB expansion has already occured, so the inclusion of 
the term with  $\partial v_{\parallel}/\partial \theta$ into (\ref{m_si})
does not change the final results 
 significantly. Moreover, if we do include it, 
it would only suppress sliding, as can be seen from 
(\ref{motion}), so that our results for sliding and 
shell expansion can be considered to give upper bounds for the final height
of the shell.

\section{Rayleigh-Taylor instability}

 It is tempting to expect that for even more powerful SBs, than 
 $N=100$ SNs, the
effect of sliding will be even more pronounced and the 
final size of the SB may reach 
scales comparable to the size of Galaxy, thus making possible the escape 
of the flux. But when the luminosity of the SB approaches
 $10^{38}~{\rm ergs~ s^{-1}}$, corresponding to the number of SNs 
$N\approx 100$, 
another effect becomes important for the fate of SB.
At such a large luminosity the shell starts accelerating  some time
before its expansion could be stopped by the gravity and it would accelerate,
in principle, to a very high velocity if there is enough time for it.

However, it was first noticed by Mac Low $\&$ McCray (1988) and then proven
numerically by them (Mac Low $\&$ McCray 1989) that as soon as 
the shell starts to 
accelerate it becomes Rayleigh-Taylor unstable and eventually breaks up.
They called this process a ``blowout'' of the shell into the halo. 
Indeed, the effective gravity in the moving shell in the case of 
acceleration is directed towards the center of SB, that is 
the dense cold gas in the  shell is pushed by  rarefied hot gas 
of the interior and this  leads to the instability.   
The shell fragments into blobs of cold, dense gas, which continue moving with 
velocities they attained before fragmentation. There is no further
significant 
acceleration of the shell, because the hot rarefied gas from the interior of 
the SB escapes
 into halo thus allowing the inner pressure to drop. After that time 
each blob will move ballistically, sweeping up some small mass, though this 
effect might be not so important at high altitudes, where the density of 
all the components of the ISM is very low. This means that if the speed of the 
blob at this time 
is less than the escape velocity from the Galaxy, matter can not
 leave but must return to the Galactic plane from halo and  there is 
 no contribution to the flux escape.
For this instability  the galactic gravity
contributes to the
effective gravity as well, thus making the shell unstable even when it is still
decelerating.

\begin{figure}
\vspace{16.0cm}
\includegraphics{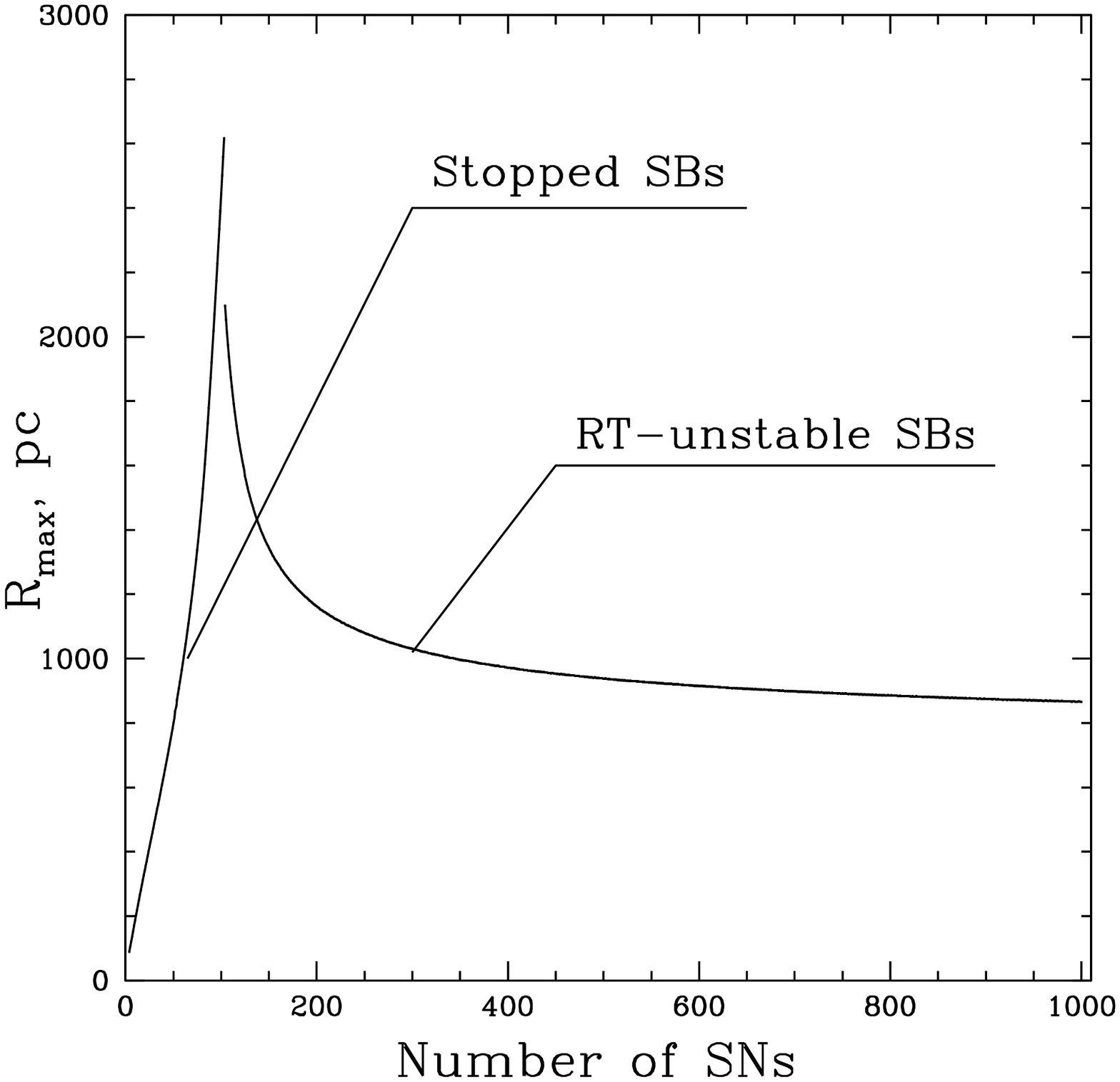}
\includegraphics{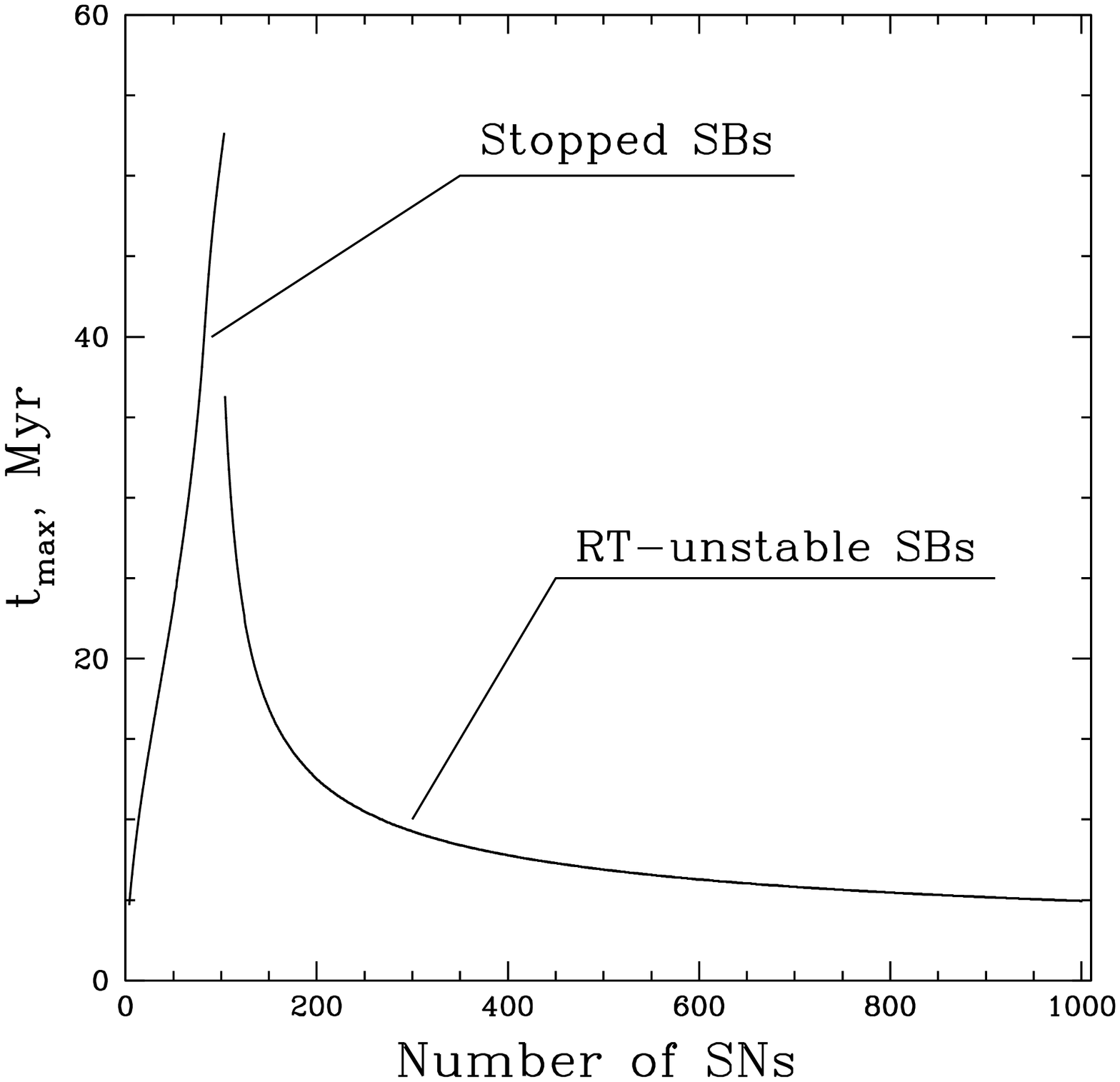}
\caption{
(a) Final size of the SB, $R_{max}$, as a function of $N$ -- the
 number of SNs producing
different luminosities. Shells of the low luminosity bubbles
are  stopped by galactic gravity (left branch), while those of the  
more powerful bubbles (right branch) are disrupted by the Rayleigh-Taylor
instability.
(b) Distribution of times when the SB expansion stops, $t_{max}$,
either due to the 
galactic gravity (left branch) or the Rayleigh-Taylor instability 
(right branch)
for different
number of SNs.}
\label{rad-time}
\end{figure}

In our situation, when we have a semi-infinite hot rarefied medium 
and a dense slab of gas of finite thickness $H$ up on it, the fastest
growing mode of the instability has a scale of the order of $H$, so that
 the growth rate of the instability is given by
\begin{equation}
\gamma^2=-\left(\ddot R+g\right)/H. 
\end{equation} 
The increment of instability or the amount of e-foldings
reduces to 
\begin{equation}
\int\gamma dt=\int dt\sqrt{\left(g+\ddot R\right)/H}.
\end{equation} 
For the fragmentation to proceed effectively we require 
\begin{equation}
\int\gamma dt > 1.
\label{incr}
\end{equation}

To get the shell thickness, let us note that the gas initially located
between the heights 
$z_0+z$ and  $z_0+z+dz$ is deposited into the shell between
$h+dh$ and $h$ from the shell outer surface. The conservation of 
the number of particles accounting for the sliding gives that
number density of particles in the shell at local thickness $h$
related to the number density of particles initially at the point 
$z_0+z$ as
\begin{equation}
\frac{\sigma}{\sigma_0}n_0(z_0+z)z^2 dz=n(h)R^2 dh, 
\end{equation}
where 
\begin{equation}
\sigma_0(t)=\frac{1}{R^2(t)}
\int\limits^{t}_{0}R^2 \dot R \rho(t^{\prime})
dt^{\prime},
\label{sigma_0}
\end{equation}
is just $\sigma$ without sliding. The number density $n$ is given by
$n=P_{sh}(h)/kT_{sh}$, where $T_{sh}$ is the temperature in the shell,
which we assume to be equal to $10^4$ K, and $P_{sh}$ is the pressure
at the local thickness $h$ in the shell. Here we have taken into account
only the thermal pressure and neglected the cosmic ray pressure.
This seems to be the reasonable assumption, because, as will be shown later,
 they have very high drift velocity and may easily escape from  
the compressed shell along the 
magnetic field lines deposited into it, since the lines 
themselves leave the shell.
 The pressure $P_{sh}$ is comparable
to the inner pressure $P_{in}$, because inside the shell it has to drop
from $P_{in}$ on the inner surface  to $P_{out}\ll P_{in}$
outside the shell.

Combining all these considerations  we obtain  that 
\begin{equation}
H\approx\frac{\sigma}{\sigma_0}
\int\limits^R_0 \frac{n_0(z_0+z)k T_{sh} z^2}{P_{in} R^2}dz.
\end{equation} 

Numerical estimates show that during the first several 
e-foldings of  the Rayleigh-Taylor instability, 
the shell thickness $H$ changes very little, 
less than $10\%$. During this time the size of the shell
changes less than $\sim 30\%$, so that the 
geometrical effect of the
stretching the scale of the perturbation mode
 in the expanding shell is quite moderate.
For this reason the modes which were unstable at the very onset of the
 instability stay unstable during several e-foldings thus developing 
the nonlinear stage of the instability and disrupting the shell.

During the development of the instability
the  velocity of the shell does not change drastically,
it is larger only $(3-5)\%$ than 
the velocity at the very onset of instability.
Thus, we may safely assume that after the  Rayleigh-Taylor instability is 
fully developed in the shell, we get no further acceleration of the shell 
fragments.

We consider the role of the Rayleigh-Taylor
instability  on the fate of SBs of various 
luminosities with inner 
pressure from (\ref{pressure}) located at the altitude $z_0=100$ pc near the 
Sun. The results for maximum size of the shell and time when it either 
stops due to gravity or fragments because of Rayleigh-Taylor
instability, are shown on Figure \ref{rad-time}. 
Equation (\ref{incr}) is chosen 
to be the condition for Rayleigh-Taylor fragmentation of the shell, 
because it corresponds to the onset of the nonlinear stage of this instability
when the shell is being disrupted.
The two branches of Figure \ref{rad-time} 
correspond to the bubbles which were stopped (left
branch) and to those which were disrupted (right branch).
We see that for a chosen dependence of 
inner pressure upon time the transition to the Rayleigh-Taylor regime occurs
for $N=104$ supernovae in the SB. The maximum possible size and the greatest 
lifetime of the SB  are achieved just before  this $N$ and are equal to 
$R_{max}=2616$ pc
and $t_{max}=52.6$ Myr. After that size
is reached the  expansion drops rapidly 
due to the early onset of the Rayleigh-Taylor instability in the shell, and
we see that for SBs with $N\sim 1000$ shell fragmentation occurs at a
very early time $\sim 5-6$ Myr. It must be emphasized that at such an early
stages the approximation of constant luminosity may not be justified and 
dynamics  might be more complex. We believe  that the 
result of the analysis is not very sensitive to the model assumptions.

\begin{figure}
\vspace{8.0cm}
\includegraphics{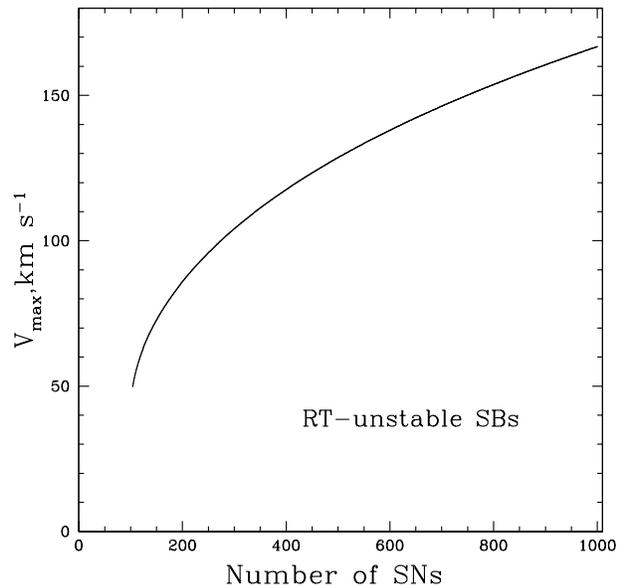}
\caption{
Velocity of blobs formed in the process of the fragmentation of
SB shell, $v_{max}$,
 versus the luminosity of the  SB (number of SNs in the initial star cluster).
Though this velocity increases with the luminosity of the SB, it does not
reach the escape velocity from the Galaxy. }
\label{vel}
\end{figure}

In Figure \ref{vel} we plot the dependence of the blob velocity upon the 
luminosity of the SB. One can see that it is sufficiently less 
than $\sim 430~ {\rm km~ s^{-1}}$, the 
estimated lower bound on the Galactic escape velocity 
(Leonard $\&$ Tremaine 1990),
so that we may conclude that shells 
formed by SBs with number of SNs in them $N\le 10^3$ do not give rise 
to the mass and flux 
outflow from the Galaxy. Even if we go to a SB with a luminosity
an order of magnitude larger ($N=10^4$), the velocity at the moment of 
fragmentation is only 
$v_{max}\approx 336~{\rm  km~ s^{-1}}$, 
 which is obviously not enough to leave the Galaxy.

\section{Importance of the shape of the SB top}

In  our treatment of the SB expansion we 
have considered the shape of the shell 
near its top to be spherical. This enables us to use a simple fact that 
in this case the projection of Galactic gravitational acceleration along the 
shell is just $g_{\parallel}=g\sin\theta$ which sufficiently simplifies 
the problem.  

In reality, of course, the surface density is not uniform on the 
top but depends upon the angle $\theta$. The
nonuniformity grows with growing $\theta$.
The inner pressure will accelerate parts of the shell closer
 to the 
top stronger than ones further from the top and it will distort the form of 
the shell. This distortion in its turn changes $g_{\parallel}$ which
influences the sliding of the matter and thus leads to further changes 
of the shell's form.
The accurate treatment of the problem requires including this
effect self-consistently in our calculations, but we can avoid this by noting 
that influence of a change of the shell's shape on the process of sliding
can be attributed to the change in gravitational acceleration $g$, rather
than the projection angle.  
     
We carried out calculations identical to those with spherical top but 
have taken $g$ in equation (\ref{kappa-s}) to be $4$ times larger 
than it is in reality. The result was that, as the luminosity of the SB 
was increased,
they expanded faster, due to the more effective sliding of the matter
from the top. But this in turn lead to a more rapid onset of Rayleigh-Taylor  
instability in the shell: it started to develop  for SBs 
with  $N>69$. The bubble with $N=69$ 
 stops at a time $t_{RT}=45.7$ Myr and size
$R_{max}=2398$ pc. The conclusion is obvious: the change of the 
shell shape might influence the sliding of the matter, but it leads to 
 the development of the  Rayleigh-Taylor  instability for even 
 smaller luminosities than in the case of the bubbles with the 
spherical top, 
making the  impossibility of the expulsion of the flux 
from Galactic disc even more certain. 
That is why we think a more self-consistent approach
to the problem of the shell shape will not change the general result.

\section{Different pressure law}

The pressure law (\ref{pressure})
which we used in all our calculations was derived actually for the 
case of the SB expanding in a uniform medium 
and thus may be not a very good approximation 
for our purposes especially when the sliding of the matter influences
the expansion of the shell and its size cannot be described by equation
(\ref{R}). For that reason we decided to use a different, more 
realistic pressure 
dependence to check if it makes a 
significant difference in our results.

Maciejewski $\&$ Cox (1999) proposed a simple, explicit, analytical 
approximation for the kinematics of the blast wave propagating in 
an exponentially stratified medium: 
\begin{equation}
\rho_0=\rho_{\star}e^{-z/h},
\end{equation} 
where $\rho_{\star}$ is the density at the explosion site and $h$
is the stratification length scale. They considered an explosion in 
the framework of the Kompaneets approximation (Kompaneets 1960) 
with inner pressure constant
throughout the volume engulfed by the shock. They showed that the form of 
the shell is very close to an ellipsoid with minor and major axes
$b$ and $a$  related by
\begin{equation}
\tan\frac{b}{2h}=\sinh\frac{a}{2h}
\label{minor}
\end{equation}
and the distance from the explosion site to the ellipsoid center, $s$,
\begin{equation}
\tan\frac{s}{2h}=\cosh\frac{a}{2h}.
\label{displ}
\end{equation}

\begin{figure}
\vspace{16.0cm}
\includegraphics{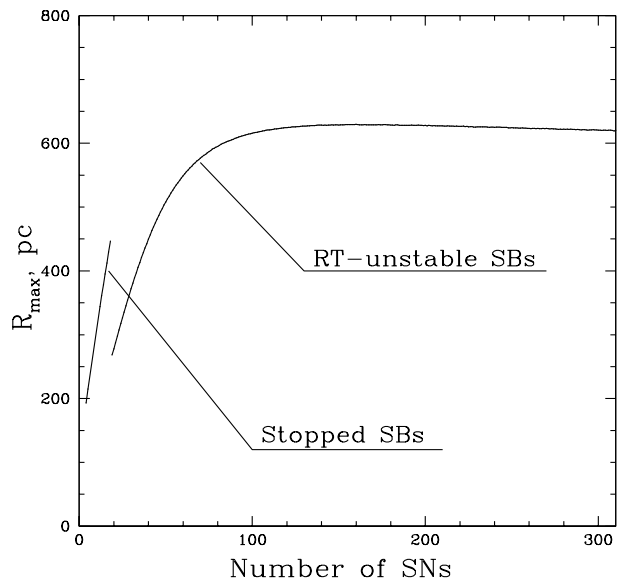}
\includegraphics{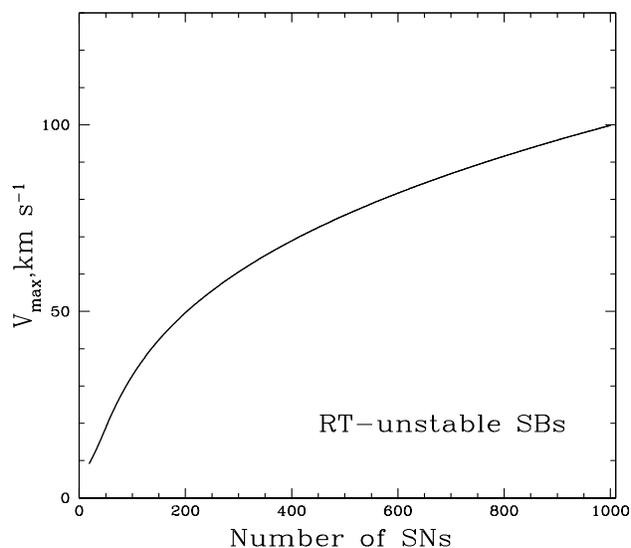}
\caption{
(a) Final size of the shell, $R_{max}$, and
(b) velocity of the shell, 
$v_{max}$, at the moment of stopping or fragmentation 
versus the number of SNs in SB. Pressure  law is different from 
the case represented on the Figure \ref{rad-time} and is given by
equations (\ref{p_in}),(\ref{relat}), and (\ref{minor}).}
\label{rad-vel}
\end{figure}

At the same time, Weaver et al (1977) showed that internal energy of the
SB interior is constant fraction of the total energy and for a
spherical SB expanding in a homogeneous medium
\begin{equation}
E_{in}=\frac{5}{11}L_{SN} t.
\end{equation}
We assume that this is also valid for the case of SB in 
a nonhomogeneous density 
 distribution, so that
 that inner pressure 
\begin{equation}
P_{in}=\frac{E_{in}}{V}=\frac{5}{11}\frac{L_{SN} t}{\pi a b^2}.
\label{p_in}
\end{equation}
We relate the semi-major axis $a$ to the distance from the center of explosion to
the top of the shell: $a+s=R$, by
\begin{equation}
\frac{a}{2h}+\log\left(\cosh\frac{a}{2h}\right)=\frac{R}{2h}.
\label{relat}
\end{equation}
Formulae (\ref{p_in}),(\ref{relat}), and (\ref{minor}) 
give us $P_{in}$ for a given $R$. The approach of Maciejewski $\&$ Cox
(1999) includes neither galactic gravity nor the slippage from the 
top but 
it takes the inhomogeneity of the surrounding medium into account
and enables us to test the stability of our results against different
model assumptions.  

The real ISM contains many components distributed with various length scales
so we take rather arbitrarily $h=200$ pc in our case. 
The particular choice of $h$ turns out not to play a significant role. 
The results for superbubbles of various luminosities 
located at $z_0=200$ pc near the Sun
are shown in Figures \ref{rad-vel}a and \ref{rad-vel}b. 

We see that differences are only quantitative compared to the case of 
pressure law (\ref{pressure}). Strong Rayleigh-Taylor instability starts 
to dominate the kinematics of the shell when the number of SNs in SB is larger
than $N=18$. This SB reaches the maximum size $R_{max}=447$ pc at a time 
$t_{max}=41$ Myr.  Due to the specifics of the the chosen pressure law,
SB with the higher luminosity, developing the Rayleigh-Taylor instability,
 reach somewhat larger size, maximum is $R_{RT}=630$ pc. 
If we take, for example, $h=100$ pc, then the shells
start to be  not stopped by the gravity but disrupted by Rayleigh-Taylor 
instability
even for smaller $N$.
Thus, again, for powerful SBs,
expansion is limited by the shell fragmentation so that all major
results of the consideration with the simplified pressure law remain valid.

\section{Possible importance of the CR pressure in the shell}

Kulsrud (1999) proposed that sliding of the matter from the top of the 
SB shell 
may be inhibited to some extent by cosmic ray (CR) pressure gradient, 
thus further
supporting our conclusion about the impossibility of flux expulsion from 
the Galaxy. Now, on the basis of the better knowledge of the processes 
going on in SB shell, we can check this idea in more detail.

   Let us consider the magnetic flux tube with the cross section 
constant along the tube
in the shell which reached the size $R$ and expands with velocity $u=\dot R$.
 This assumption is good for the 
cylindrical shell, with its axis lying in the Galactic plane, but it 
will be clear further that this assumption is not important.
The continuity equation for the CR along the flux tube reads:
\begin{equation}
\frac{\partial }{\partial t}\left(n_{CR} R\right)+
\frac{\partial}{\partial \theta}
\left[n_{CR}\left(v_{\parallel}+v_d\right)\right]=0,
\label{cont1}
\end{equation}
where $n_{CR}$ is the number density of the CR,
 $v_{\parallel}$ and $v_d$ are the velocities of matter along the shell
 with respect to the 
rest system  and of CR with respect to the matter correspondingly.
The drift of the CR along the magnetic field is determined by the gradient 
of their number density and by the scattering of the CR by the Alfv$\acute{\rm e}$nic 
turbulence. The scattering of CR by the self-generated Alfv$\acute{\rm e}$n waves 
was first considered by Kulsrud $\&$ Pearce (1969) and Wentzel (1969)
and they showed that CR cause the Alfv$\acute{\rm e}$n waves to grow with a rate
\begin{equation}
\Gamma=C\frac{\pi}{4}\Omega_0\frac{v_d-v_A}{v_A}\frac{n_{CR}(\epsilon)}{n},
\label{Gamma}
\end{equation}
where $\Omega_0$ is the nonrelativistic cyclotron frequency, 
$n_{CR}(\epsilon)$ is the number density of CR with energy greater than
$\epsilon$, $n$ is the density of the ISM, $v_A=B/\sqrt{4\pi n m_H}$ is 
the Alfv$\acute{\rm e}$n speed ($m_H$ is the mass of hydrogen atom), 
and $C$  is a
 constant of the order of unity whose value depends upon the energy 
spectrum of CR. These waves are also  nonlinearly
damped by the scattering 
of the beat waves on the ions:
\begin{equation}
\gamma_d= \sqrt{\frac{\pi}{8}}\frac{v_T}{c}
\Omega\left(\frac{\delta B}{B}\right)^2,
\label{damp}
\end{equation}
(Lee $\&$ V\"olk 1973; Kulsrud 1978)
with $\delta B$ being the magnetic field perturbations due to the Alfv$\acute{\rm e}$n waves,
$\Omega$ -- relativistic Larmor frequency,
and $v_T$ -- thermal velocity of ions in ISM. Scattering of CR occurs at a rate
\begin{equation}
\nu=\Omega\left(\frac{\delta B}{B}\right)^2,
\end{equation}
(Kulsrud 1995),
so that the typical mean free path of the CR is 
\begin{equation}
\lambda=\frac{c}{\nu}=\frac{c}{\Omega}\left(\frac{B}{\delta B}\right)^2.
\label{mfp}
\end{equation}

On the other hand, the drift velocity is given by
\begin{equation}
n_{CR} v_d=c\frac{\lambda}{R}\frac{\partial n_{CR}}{\partial \theta}.
\label{drift}
\end{equation} 
Assuming that $v_d\gg v_A$, that Alfv$\acute{\rm e}$n waves are in equilibrium,
$\Gamma=\gamma_d$,
 and combining equations (\ref{Gamma}),(\ref{damp}),
(\ref{mfp}), and (\ref{drift}) we get
\begin{eqnarray}
|v_d|=\left(\sqrt{\frac{2}{\pi}}\frac{c v_T n v_A}{C \Omega_0 n_{CR}^2}
\left|\frac{1}{R}\frac{\partial n_{CR}}
{\partial \theta}\right|\right)^{1/2}
\nonumber\\
=\left(A
\left|\frac{1}{R n_{CR}^2}\frac{\partial n_{CR}}
{\partial \theta}\right|\right)^{1/2}.
\label{drift_f}
\end{eqnarray}

If we
introduce the characteristic drift velocity
\begin{equation}
v_D=\sqrt{\frac{A}{n_{CR} R}},
\label{v_drift}
\end{equation}
and suppose that 
\begin{equation}
v_D\gg v_{\parallel},
\label{appr}
\end{equation}
then it is easy to see from equation (\ref{cont1})
that when $v_d\approx v_{\parallel}$
the variation of the 
 CR density can be expressed as
\begin{equation}
n_{CR}R=n_{CR_0}R_0\left(1+\left(\frac{v_{\parallel}}
{v_D}\right)^2\chi_{1}(\theta)\right),
\end{equation}
where $\chi_{1}(\theta)$ is some function of order of unity.

\begin{figure}
\vspace{8.0cm}
\includegraphics{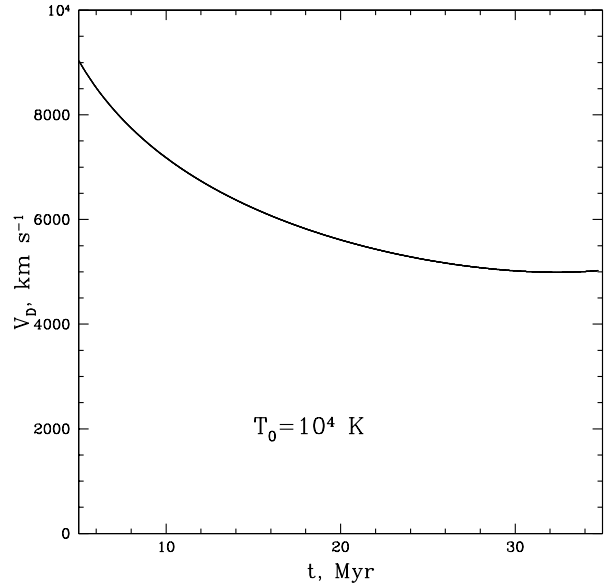}
\caption{
Characteristic drift velocity 
of CR, $v_D$, in the shell of SB due to the scattering of the CR
by the self-generated Alfv$\acute{\rm e}$n turbulence. Note the significant difference 
in the magnitudes of $v_D$ and the sliding velocity $v_{\parallel}$
in the shell
(Figure \ref{velocities}).}
\label{fig-drift}
\end{figure}

To see if the condition (\ref{appr}) is fulfilled, we calculated 
$v_D$ given by (\ref{v_drift}) and (\ref{drift_f})
at various time moments for the SB with $N=75$ SNs,
taking into account the compression of the matter and CR in the shell,
which we calculate in the manner similar to the calculation of the thickness
of the shell, and
supposing that the temperature in the shell after cooling is $T_s=10^4$ K.
Compression of matter in the shell (and, consequently, of the CR) depends 
upon the temperature of ambient ISM $T_0$, so  that
$v_D$ scales as $v_D \propto T_0^{-3/4}$. In Figure \ref{fig-drift} we
plot $v_D$, defined by (\ref{v_drift}) for $T_0= 10^4$ K.

One can see from comparison of Figures \ref{fig-drift} 
and \ref{velocities}, 
that condition (\ref{appr}) is always fulfilled for $T_0=10^4$ K. 
It might be violated if $T_0\sim 10^6-10^7$ K, that is
if the SB expands into the predominantly hot ISM component, but 
near the top the condition (\ref{appr}) is always fulfilled
even for these high temperatures, because 
$v_{\parallel}\propto\sin^2\theta$.

In the approximation given by (\ref{appr}) the spatial CR 
density perturbations
are small and the first term in equation (\ref{cont1}) is negligible.
This means that the drift velocity is almost equal to the 
velocity of matter $v_{\parallel}$
but opposite to it in the direction, so that CR slide through the 
matter to maintain constant spatial density and the time variations 
of their density are only due to the shell expansion. Then the equation of 
continuity reduces to
\begin{equation}
\frac{\partial}{\partial \theta}\left(n_{CR}v_{\parallel}-
\sqrt{\frac{A}{n_{CR}}\frac{\partial n_{CR}}{\partial \theta}}\right)=0.
\end{equation}
This equation can be integrated with initial condition 
$\partial n_{CR}/\partial \theta=0$ at $\theta=0$, where $v_{\parallel}=0$, 
to give
\begin{equation}
\frac{\partial n_{CR}}{\partial \theta}=n_{CR} v_{\parallel}^2
\left(\frac{A}{n_{CR}R}\right)^{-1}.
\end{equation}

Then the ratio of CR pressure term in the equation (\ref{motion}) to the 
gravitational acceleration along the shell is
\begin{eqnarray}
\frac{\nabla p_{CR}}{\rho_s g_{\parallel}}=\frac{\partial p_{CR}}
{\partial n_{CR}}\frac{\partial n_{CR}}{\partial \theta}
\frac{1}{R g\sin\theta\rho_s} \nonumber\\
\sim \frac{p_{CR}}{\rho_s g R}
\left(\frac{v_{\parallel}}{\sin\theta v_D}\right)^2 \sin \theta\sim
10^{-4}\sin\theta,
\end{eqnarray}
for $T_0=10^4$ K, or maybe $\sim 0.1\sin\theta$ at $T_0\sim 10^7$ K,
so that the effect the CR pressure gradient has on the sliding of the 
matter from the shell's top is negligible for all interesting cases, and, 
actually, not only at the shell's top, but also for $\theta\sim 1$.

Obviously, the assumption that the flux tube has constant cross section 
and the cylindrical geometry implied here are not important, because 
the uniformity of the CR pressure is achieved due to the very high
characteristic drift velocity $v_D$ 
of the CR compared to the sliding velocity, which perturbs 
the uniformity, and not due to any peculiarities of the flux tube geometry.

\section{Summary}   

In this paper we consider the effect of the downsliding of the matter
which takes place in the expanding superbubbles for application to the
expulsion of the magnetic flux from the Galaxy. This expulsion 
is an important ingredient of the  $\alpha{\mbox -}\Omega$
dynamo theory. However, it is shown that even the inclusion of the sliding 
into the calculations of the kinematics of the superbubbles, 
does not enable matter and frozen in flux to leave the Galaxy in SBs.
One must note, that the impossibility of the flux escape from the Galaxy
weakens significantly the Parker's (1971) argument against the primordial
magnetic field, because there is actually no mechanism  to expel it.

Some authors (Korpi et al 1999a, Korpi et al 1999b) have considered
dynamics of the superbubbles in the gravitational field of the Galaxy
in greater detail and they also find that the matter does not reach
terminal velocities larger than the escape velocity from the Galaxy.
 In their simulations 
they do observe the development of the 
SB and its blowout from the disk, but at the height of several kiloparsecs
the velocity of the matter is too small for the matter and the field 
to leave the Galaxy, which 
agrees with our conclusions. However, they do not comment on its
relation to the dynamo.

Other authors (e.g. Hanasz \& Lesch 1998, Moss et al 1999), 
more directly concerned with dynamos, do not investigate 
the dynamics of escape
but merely assume that once the magnetic field lines reach the 
boundary of the disk, they are advected away by some mechanism 
leading to a vacuum boundary conditions. They do mention magnetic buoyancy
as a possible escape mechanism but buoyancy is unlikely 
to be important outside the disk and clearly plays no role in the SB 
escape mechanism, especially when the magnetic fields are first amplified 
from weak seed fields. 

In this paper an analytical formalism for the consideration of sliding 
was built on reasonable assumptions, which enabled
to include the back-reaction of lowering the density of the matter
on the top of the 
SB, on the expansion of the shell in the radial direction.
All the SB, depending on their luminosity and conditions in the surrounding
ISM were demonstrated to fall into two classes: low luminosity SBs
 which are  stopped by the Galactic gravitational field and fall back, and
powerful SBs, which are possibly 
able to reach low density regions at high altitude.
Even without sliding these powerful SBs might be able to
expel matter from the Galaxy.
However, before this happens
the shells of these high luminosity  SBs fragment into
separate blobs due to the Rayleigh-Taylor instability, which develop 
when the shells start accelerating at the high altitudes, where the
density of the matter and the outer pressure are very small. 
These separate fragments of the shells continue to
move ballistically with velocities too small to leave the Galaxy.

It was shown that our conclusion about the
impossibility of flux to escape in SB explosions 
does not depend essentially
upon the details of the inner pressure behavior in 
the SB or the shape of the SB top. The inclusion of the CR pressure 
gradient in the shell also does not influence the downsliding because
of the high uniformity of CR density in the shell caused by the very 
large diffusion velocity of CR in the shell.

It should be remarked that throughout the paper we have assumed that
magnetic field lines are tied to the gas. If the ionization is the shell
is low, then ambipolar diffusion (under the influence of the CR pressure
gradient perpendicular to the shell) could allow some small slippage of the 
field lines through the neutral component and, thus, some small amount of 
escape. We do not discuss this possibility in this paper.

All this lead us to the conclusion that if there does exist
some mechanism responsible for the flux expulsions from the Galaxy 
needed by the $\alpha{\mbox -}\Omega$ dynamo
this mechanism is not superbubbles. The question of
whether such a mechanism exist remain an open one.

\section{Acknowledgements} 

One of the authors (RRR)
would like to acknowledge the financial support of this work by
the Princeton University Science Fellowship.

\end{document}